\documentstyle[epsfig,graphicx,psfrag]{article}                
\begin{document}
\begin{center}
\Huge
{\bf Study of changes in the Solar Environment}

\LARGE
{M.N. Vahia$^1$}\\
\Large
\noindent $^1$ Tata Institute of Fundamental Research, Homi Bhabha Road, Colaba, Mumbai 400 005, INDIA email: vahia@tifr.res.in\\
\end{center}
\bigskip

\centerline{\bf Abstract}
Recent studies have shown that the local environment of the Sun is a complex one with the presence of several supernova shock bubbles and interstellar clouds. Even within this environment, the Sun is moving with a velocity of about 15 km/s toward the interior of the galaxy. We examine the changes in the local ISM environment experienced by the Sun in the past and then discuss the interaction of the Sun with these different environments. Lastly, we enquire how this must have changed the cosmic ray environment of the earth since this remains one of the few methods to test the changes in the local ISM environment over astronomical time scales.

\medskip

\vskip 0.2in

\section{Introduction}

Recent studies of the local interstellar medium reveal that the Sun is in a complex environment where at least three and probably more bubbles or cavities of probable supernova origin are interacting (see e.g. Breitschwerdt, 2001, Frisch, 1995, 1998, Maiz-Apellaniz, 2001, Smith and Cox, 2001, and references therein). 

Reviews by Frisch (1998, 1995) have mapped the local interstellar matter (LISM) from several observations based on back scattering and extinction of soft X-rays coming from nearby sources. These maps show that the local interstellar medium is largely governed by a collection of at least three large bubbles or super bubbles that seem to be shells of supernova remnants of age between 5 and 10 Million years. A large bubble in which the Sun seems to be sitting is referred to as local bubble. These observations have been re-enforced by radio observations of more recent studies based on the measurements of the turbulence in the nearby pulsar radio emission due to the LISM plasma (Ramesh Bhat and Gupta, 2002, Ramesh Bhat, Gupta and Pramesh Rao, 2001). They have modeled the inferred plasma distribution into a 3 component model that gives a description of the LISM derived from X-ray observations.  
Redfield and Linsky (2002) have attempted to study the UV observations of FeII, MgII and CaII to determine the structure of the LISM up to 100 pc. they have shown that combined studies of absorption features and Doppler shifts indicate a fairly non uniform distribution of the LISM. Further high resolution studies in ultraviolet (Shelton, 2002) and infrared bands (Franco, 2002) from select regions show that there is evidence of hot but quiescent plasma in the LISM. A more detailed compilation of the H$_\alpha$ lines in the Milky Way by Haffner (2002) has shown that conditions of the warm interstellar medium and diffuse ionised gas suggest a rich mix of filamentary structures in the ISM. Similarly, first results from the FUSE mission (Moos et al., 2002) suggest that UV emission in the the warm component of the local bubble is homogeneous indicating an old bubble.

Some of these shells seem to be disintegrating and forming fluffy clouds that are drifting within these shells or loops (see also Ricardo and Beckman, 2001; Seth and Linsky, 2001; Smoker et al., 2002). Meisel, Diego and Mathews (2002) have calculated the detailed trajectories of several dust grains approaching the sun and shown that they are indeed of interstellar origin. The dust content of the bubble have been estimated from the Ulysses by Ingrid and Hirshi (2002). 

These observations are in good agreement of models with the ISM that involve several shells in dynamic equilibrium between slowly dying supernova remnants of age of the order of a few million years (e.g. Smith and Cox, 2001) that are slowly disintegrating into small clouds.

Donato, Manrin and Taillet (2002) have attempted to model the confinement of cosmic rays in these bubbles and the production of radio active nucleids in spallation inside the local bubble. They have shown that production rates of some radioactive nuclei such as $^{10}$Be, $^{28}$Al and $^{36}$Cl can be important markers of the spallation rates in the LISM. 

It is therefore clear that we have a fairly consistent broad picture of the local interstellar medium (see figure 5, Frisch 1995). Another interesting feature is that the Sun itself is not at rest in the local rest frame (Frisch, 1995, 1998) and has a proper motion of about 15 km/s toward the inner part of the galaxy. Dehnen and Binney (1998) have given a more detailed value of the movement vector of the sun in the local ISM using the Hipparcos data. They show that the best fit vector for the sun is $u_0=10.00\pm0.36~km/s$, $v_0=5.25\pm 0.62~km/s$ in the plane of the Galaxy and $w_0=7.17 \pm 0.38 km/s$ perpendicular to the plane of the Galaxy towards the galactic plane. However, this motion in the perpendicular plane is believed to be cyclic with a period of 66 million years (see also Frisch, 1998). The result is that the sun has been encountering a large range of ISM conditions over the last few million years. While attempts have been made to model the dynamics of the ISM no studies have been undertaken to understand the interaction of the sun under these varying conditions. In the next section we consider the changing environment seen by the Sun over the last few million years. In section 3 we consider various processes by which the sun interacts with the local interstellar medium and in section 4 we concentrate on the modulation of the average galactic cosmic ray fluxes due to these differing environments and show that the observed cosmic ray fluxes on earth would vary significantly due to different ambient LISM conditions. 

\section{Position and relative movement of the Sun in the Galaxy}

We begin backtracking the path of the sun based on our current understanding of the velocity of the sun in LISM and the LISM environment. For this we refer to figure 5 of Frisch (1995). The sun is moving in the LISM toward the region of Galactic coordinates of $(l,\beta$) = (51$^o$, 23$^o$)with a speed of 15.4 km s$^{-1}$) (Frisch, 1995)with respect to the local rest frame. The Scorpius-Centaurus association is a large number of massive stars. Several supernova explosions have occurred in this region over the last 15 million years. It is generally believed that the Sun is inside the supernova bubble created by supernova remnants from this region. Smith and Cox (2001) consider the option that the local environment is a result of two or three supernovae that went off in the region occupied by the Local Bubble that itself could have originated from the Scorpius Centaurus Association. The center of the Gould belt of young stars is located at a mean galactic longitude of 110$^{o}$ with a diameter of 200 pc (Poppel, 1997). The Taurus and Perseus region lies in the anti galactic center region while the Orion loop lies in the Galactic Longitude of $l$ about 200$^o$. 

\begin{table*}
\small
\caption{Solar movement in the ISM }
\label{SMtable}
\footnotesize
\begin{tabular}{lccccl} \\

\hline
\multicolumn{2}{c}{\underline{Time (yBP)(a)}}&	dT	& Region&	Den. (cm$^{-3}$)&		Comments(b)\\
 Start	& End &	(y)& 	 &(Temp (K))	&	\\
\hline						
0	& 1.0 10$^4$ & 1.0 10$^4$ &	fluff & 	0.08	&	Size 5 pc, From SCA,  \\
&&&&~(7000)&moving in  orthogonal direction to Sun 20 km/s  \\
&&&&& B= 1.5 $\mu$G. Highly Inhomogeneous (1\\
1.0 10$^5$ (a)&	3.1 10$^5$	& 6.0 10$^3$ &	SN(?)&	4 (104)	&	Geminga? (2,3) \\

1.0 10$^4$ &	1.0 10$^5$ & (0.1-1)10$^5$ &	bubble &	0.04 ~(106 )	&	Expanding bubble from SCA. Age 15 My \\
&&&&& vel $\approx$ 4 10$^4$ km s$^{-1}$ orthogonal to Sun B~ 1.5 $\mu$ G  \\
&&&&& crossed the Sun 250,000to 400,000 y  ago (1,3,4,5)\\

 1.0 10$^5$ & 1.0 10$^7$	& (0.01-1) 10$^7$	& Inter-arm &	0.0005 (~100)	&	(1,6) \\

1.0 10$^7$(a)&	2.5 10$^7$& 3.0 10$^6$ & HI shock &	4 (10$^4$ )	&	From Gould belt (7). Duration is taken as that for   \\
&&&&& the earlier SN.(3,6)\\

2.5 10$^7$(a) &	4.0 10$^7$ &	5.0 10$^5$	& Orion belt	& 10$^4$ (~10)&		Size ~ 5pc, Sun must have  5 10$^5$taken yrs  to cross \\
&&&&& at current velocity (7,8)\\
4.0 10$^7$ &	5.0 10$^7$	& (2.5-4)10$^7$ &	inter-arm &	0.005 (10) & 		(6) \\
5.0 10$^7$ &	6.0 10$^7$	& (4-5)10$^7$	& Persius arm	& 1 (1000)	&	Duration uncertain(6) \\
$\leq$ 6 10$^7$	&&&&&					(9) \\
\hline
\end{tabular}
\smallskip
\footnotesize

(a) For SN the start and end time are the interval between which the SN spike crossed the earth	        (b) Bracket numbers in comments refer to notes. \\
References:	1 Frisch (1995);		2 Ramadurai(1993); 3 Sabalska  et al(1991)	4 Egger et al(1996) 5 Van der Walt and Wolfendale (1988) 6 Bash (1987) 7 Clube and Nupier (1984)

\bigskip
Notes to the table:

1) SCA refers to Scorpius Centaurus association at l = ~ 300-360$^o$. About 35 SNR have been identified in this region (Whiteoak and Green, 1996).

2) While Frisch (1995) claims that the evidence for Geminga is not there, Ramadurai(1993) has claimed the passage of a SN shock around 30,000 yrs ago that arose from a source around the age of Geminga. The sun crossed the vicinity of the parent star of Geminga 10$^7$ yrs BP while the SN itself is 3 10$^5$ yrs old (Pavlov, 1997). The temperature and density are typical values for SNR (Harwit, 1998). The shock passage is calculated assuming a shock front velocity of 1000 km/s (1 pc/My) and taking an effective thickness of 1 pc (Harwit, 1998). These parameters are similar to those derived by Szabelska, Szabelski and Wolfendale (1991). Based on $^{10}$Be data they argue for the passage of a supernova shock front near the Sun about 30,000 yrs ago. 

3) Another SN is just arriving from the SCA association, age \~ 11 My. 

4) The bubble is from the SN shell from the Upper Centaurus Lupus of SCO. Age \~ 14-15 My BP.

5)  The inter-arm region is derived from figure 5 of Bertrische (1998) and Bash (1987), figure 1. 

6) van der Walt and Wolfendale (1998) argue that the gamma ray data suggest that the inter arm cosmic ray are of lower intensity and have a steeper spectrum.

7) More than 130 weak line T Tauri (O type) stars have recently been cataloged in this region (for reference see Wichmann et al., 1997)

8) Based on extrapolation as defined in note 5.

9) 
A)	We are now in the Orion structure in the inter arm region between Persius and Carina arms

B)	This is uncertain since collisions with individual objects are ignored (Bash, 1987). Clube and Napier, (1984) suggest that in the last 4.5 10$^9$ years, the sun has encountered about 10 molecular clouds. Napier (1985) has done a more detailed analysis to show that the Sun has had near collisions (impact parameter $<$ 20 pc) with 56 GMCs having $>$ 3 10$^3$ M$_{sun}$ and 8.2 close encounters with GMCs having M $>$ 10$^5$ M$_{sun}$

C)	Our galaxy is a highly evolved spiral with about 10\% of its mass in diffuse gas; the rest is in stars. The sun is located slightly above the galactic plane, and is moving through space at 16.5 km/s with respect to the local standard of rest. The Sun oscillates in the galactic plane with a period of 33 My (i.e. it goes $\pm$ 56 pc).
 
D) Taking the Sun's relative velocity of 16.5 km s$^{-1}$ (2 pc/My), in the local rest frame, the sun has traveled 9 kpc in 4.5 billion years. Taking the solar distance as 8 kpc from the center, this gives a radius of \~ 50 kpc. The sun therefore has traveled only about 18 percent of the distance in the local rest frame. In the galactic frame, the solar velocity is 250 km/s and hence the Sun has completed about 23 rotations in its life time.

\hrule width \textwidth height 1pt 
\end{table*}

We trace the motion of the sun {\it{back}} from this location. This is in line with the inferences of Bash (1986) who calculated the approximate path of the Sun over a long period. The details of the movement in table \ref{SMtable}. The first column of the table gives the epoch before present when the Sun was in the neighborhood of a particular region and the duration for which it was in a specific neighborhood. It should be noted that there is a broad range of period when the Sun was in the inter arm region passing by the Geminga Pulsar region and it was {\it{inside}} the Orion Nebula about 25 million years ago for about half a million years. It may be noted that the linear interpolation of the figure 5 of Frisch (1995) would indicate the Sun missing the Orion arm but the arm itself has a motion orthogonal to the solar direction with a velocity of about 1 km/s (Nagahama et al., 1998). It seems more than a coincidence that it is during this period, when the local ISM was clearly more dense that the Earth seems to have encountered larger mass extinction periods. Since high density LISM and interplanetary environment would increase friction and decay teh comet trajectories quite considerably. However, we do not discuss this issue here. 

 By extrapolation, we see that the sun passed from the Persius arm through the inter-arm region into the Orion belt. The sun escaped the supernova and entered the fluff in which it presently placed. The higher densities of diffuse gas seen within 50 pc of the sun between galactic longitudes 20$^{o}$ and 90$^{o}$ represent diffuse gas ablated from the parent molecular cloud complex. In the next section we discuss the physical processes through which the sun must interact with these regions before discussing the astronomical consequences.

\begin{figure}
\centering
{\includegraphics[height=2.5in,width=4in]{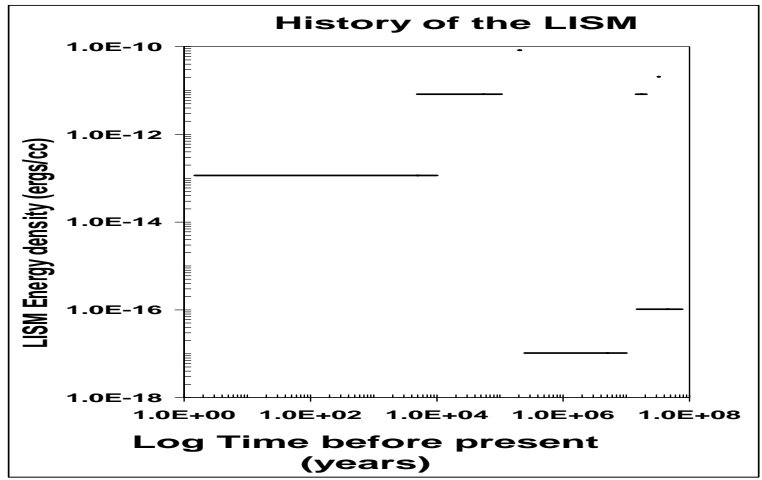}}
\caption{Energy density of the local interstellar medium as a function of time}
\label{Ld}%
\end{figure} 

In figure {\ref{Ld} we have plotted the energy density of the local ISM as a function of time. On the X-axis is the time in log scale and on Y-axis we have plotted 1.5*$k$T$_{ISM}$*$\rho_{ISM}$, where T$_{ISM}$ is the ISM temperature and $\rho_{ISM}$ is the ISM density and $k$ is the Boltzmann's constant (table {\ref{SMtable}} (see also Vahia and Lal, 2000). As can be seen from the figure, ISM densities have seen large changes in energy density over the last huudred million years for which data is given here.

\section{Physical processes of the interaction of the Sun with the local interstellar medium}

There are mainly three processes through which the interstellar gas is affected by stars (Dyson and Williams, 1997). A graphical description of these processes is given in figure \ref{pic1}.

\begin{figure*}
\begin{center}
\small
\psfrag{a}{H$_1$ n$_0$m$^{-3}$}
\psfrag{b}{H$_{II}$}
\psfrag{c}{n$_i$=n$_0$}
\psfrag{d}{Expanding Ionization Front}
\psfrag{e}{(A)}
\psfrag{f}{H$_I$ n$_o$m$^{-3}$}
\psfrag{g}{H$_I$}
\psfrag{h}{n$_i>$n$_o$}
\psfrag{i}{Shock Front}
\psfrag{j}{H$_{II}$}
\psfrag{k}{n$_i$$\le$n$_o$}
\psfrag{l}{Expanding Ionization Front}
\psfrag{m}{H$_I$n$_0$m$^{-3}$}
\psfrag{n}{H$_{II}$}
\psfrag{o}{n$_i$=n$_f$}
\psfrag{p}{Stationary Ionization Front}
\psfrag{q}{(B)}
\psfrag{r}{(C)}
\includegraphics[height=5cm]{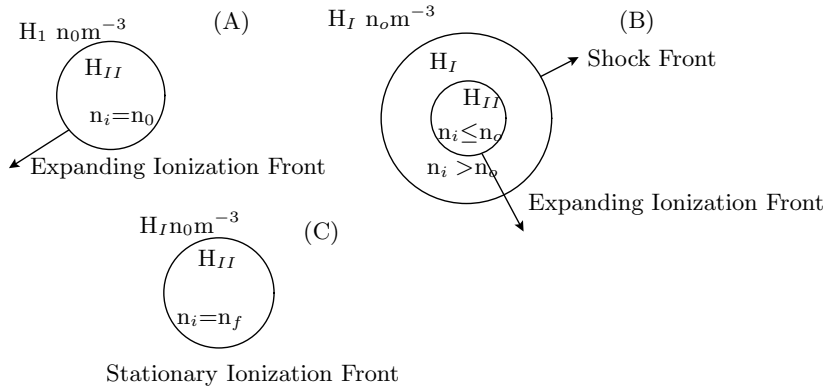}
\label{pic1}%
\caption{Evolution of the ionization induced interaction regions}
\end{center}
\end{figure*}
\normalsize

\begin{enumerate}

\item A massive hot star will photoionize the interstellar gas in its neighborhood and photons of energy $h\nu \geq I_H$ (where $(I_H)$ is the ionization potential of hydrogen) will eject electrons from the atoms. The energy excess, h$\nu$ - $I_H$, goes directly into kinetic energy of the detached electron. The protons are essentially unaffected by this process although momentum conservation demands that they experience a small recoil. The strong Coulomb interaction between protons and electrons produces the inverse process of recombination. The reversible reaction can be stated as H + h$\nu \leftrightarrow  p + e^-$. The energy of the photon produced on recombination is given by the sum of two contributions: the kinetic energy of the recombining electron and the binding energy of electron in the energy level into which it recombines. This binding energy is equal to I$_H$/n$^2$ , where n is the principal quantum number of the level into which recombination occurs. This is called the Stromgren Sphere. 

\item Due to photoionization, the gas temperature increases from about 10$^2$ K to about 10$^4$ K. The ionization process itself increases the number of gas particles and therefore the pressure by a further factor of two. The pressure in the ionized gas is thus two hundred times greater than that in surrounding neutral material. This ionized gas expands and a shock front is set into motion. The result is the formation of an outer sphere whose inside edge is defined by the ionization radius and outer radius defined by pressure equilibrium with the local interstellar medium. This is called the ionization front.

\item The other methods of interaction are via the solar wind (very high-speed continuous mass loss) which produces a mechanical piston that is bound by pressure equilibrium between the stellar wind and ISM. 
\end{enumerate}

While all three processes occur simultaneously in the interaction of the Sun with the interstellar medium, in the present study we study each process independently based on the formulation of Dyson and Williams (1997).

\subsection{Ionization front and Stromgern sphere}

In oder to analyze the ionizing interaction, we assume that the Sun was in a homogenious and isotropic medium of the given environment and the transition time scales from one interaction region to the next were much shorter than the residence time in any specific medium. While this is largely a reasonable assumption, significant acceleration of cosmic rays and higher flux at high energies can be seen when passing through a supernova shock front (see e.g. Sabalska et al., (1991). We also assume that the sun had a constant luminosity and produced Lyman continuum photons at a constant rate ($S^*$). A sphere of ionized gas is then formed around the Sun and the boundary between the ionized and neutral material moves into the surroundings at a speed, which generally approaches the speed of light. The radius of the ionized sphere is a function of time. There are two time scales of the movement of the Ionization boundary or Ionization Front (IF) and the  the surfaces of ionization and ionized gas containing the ionized gas, the Stromgren sphere.

A basic property of an IF is the velocity at which it moves. If measured with respect to the Sun, this is a fixed frame of reference. We take the gas around the Sun be at rest (in the frame of reference of the Sun). At time t, the IF is at a distance $R$ from the star and at time $t+dt$; it is at a distance $R+dR$ (figure \ref{pic1}} . We take the undisturbed neutral hydrogen density be $n_0 m^{-3}$ and let J $m^{-2}s^{-1}$ be the number of Lyman continuum photons falling normally on a unit area of IF per second. In the interior of the IF, the gas is fully ionized. Therefore, in moving the IF from $R$ to $R+dR$, enough photons must have arrived at the front to ionize all the neutral atoms lying between these positions For unit area of IF, the relationship $dt$= $n_0$ dR must be satisfied. We get

\begin{equation}
\label{eq1}
\frac{dR}{dt} = \frac{J}{n_0}
\end{equation}

The velocity given by this expression is the velocity with which the IF moves relative to the neutral gas and in the fixed frame of reference (since the gas around the IF is at rest).

	In order to relate $J$ to $S^*$, two effects need to be considered. Firstly, the IF is really a spherical surface centered on the Sun, and radiation field at the IF is diluted because of spherical geometry. Secondly, recombination takes place continuously inside the ionized region and neutral atoms are continuously being created. These atoms absorb photons traveling outwards from the Sun and cause a further reduction in the flux at the IF. If an e- and ion recombine to form an atom, a new ionizing photon will be required to separate the two particles. The photons never reach the boundary R and it will therefore not contribute to growth of the region. Ground state recombination is assumed to be balanced by an immediate ionization (called the on-the-spot assumption). Hence,
	
\begin{equation}
\label{eq2}
S\* = {4}\pi R^2 J + \left(\frac{4}{3}\right)\pi R^3 n_0^2 \beta _2(T_e)
\end{equation}

\noindent where, $\beta_2$(T$_e$), the recombination coefficient into level 2 is given by 
$$
\noindent \beta_2(T_e) = 2~10^{-6} T_e^{(-3/4)} m^3s^{-1}.
$$

Using equations \ref{eq1} and \ref{eq2}, the IF velocity is given by

\begin{equation}
\label{eq3}
\frac{dR}{dt}= \frac{S^*}{4\pi} R^2n_0 - \frac{1}{3}Rn_o\beta_2
\end{equation}

The IF velocity decreases with increasing $R$. If the Stromgren radius $R_s$ is defined as that radius at which the stellar photon output rate just balances the recombination rate in the entire ionized volume, then $S^*$ and $R_s$ are given by

\begin{equation}
\label{eq4}
S^*=\left(\frac{4}{3}\right)\pi R^3 n_0^2\beta_2(T_e)
\end{equation}

\begin{equation}
\label{eq5}
R_s=\left(\frac{3S^*}{4\pi} n_0^2\beta_2\right)^{1/3}
\end{equation}

The characteristic time, $t_R$, for which hydrogen recombines, is
\begin{equation}
\label{eq6}
t_R\approx \frac{1}{(n_o\beta_2)}
\end{equation}

 If the mass surrounding the H$_I$ region is M, the mass per unit area at the separating surface will be M/(4$\pi$ R$^2$) and the pressure inside will be 2$n_ikT$. T is the temperature of the ionized   region and the factor assumes that the number of ions n$_i$ is equal to the number of electrons. Neglecting  the small gas pressure on the outside of the sphere, we obtain the outward acceleration of the boundary as 
\begin{equation}
\label{eq7}
\frac{d^2R}{dt^2} = \frac{2n_ikT}{M/(4\pi R^2)}
\end{equation}

Integrating equation {\ref{eq6}} we get a developmental time scale of order
\begin{equation}
\label{eq8}
t_d =\left(\frac{3M}{16\pi} n_ikT \right)^{1/2}
\end{equation}

To solve Equations \ref{eq4} and \ref{eq5},we define the following dimensionless quantities.

\begin{equation}
\label{eq9}
\lambda =\frac{R}{R_s}; V_R=\frac{R_s}{t_R};\tau = \frac {t}{t_R}; \frac{d\lambda}{dt} =\frac{(dR/dt)}{V_R}
\end{equation}

Equation \ref{eq3} then becomes
\begin{equation}
\label{eq10}
\frac{d\lambda}{dt}=\frac{(1-\lambda^3)}{3\lambda^2}
\end{equation}

which has as its solution,

\begin{equation}
\label{eq11}
\lambda =(1-e^{-\tau})^{1/3}
\end{equation}

\normalsize

In the first stage, as shown in fig.\ref{pic1}, R $\le$ Rs, Equation \ref{eq10} or \ref{eq11} describes the motion.

\subsection{Implications}
We now consider the physical implications of the interactions discussed above. 
\begin{enumerate}
\item Using characteristic values only when $\tau$=4 we get $\lambda$ =0.99. Hence the equilibrium Stromgren radius is reached only after a long time. 
\item Until $R$ is very close to $R_s$, the radius of the ionized region increases at a speed much greater than the characteristic speed (the sound speed, c$_s~\approx$ 10 km/s) at which the ionized gas reacts to the sudden pressure increase caused by photo ionization. The ionized gas cannot move appreciably during this phase and the gas density consequently is unchanged by the ionization process (hence n$_i$ = n$_0$ , where n$_i$ is the ionized gas density).
\item When a time is reached greater than a few times t$_R$, the IF slows down very rapidly till its velocity is below the sound speed in the ionized gas.
\end{enumerate}

\subsection{Ionization Heating Equilibrium:}
A shock wave is set up when a piston is advanced supersonically into a gas. The expansion velocity of the ionized gas sphere (which is bounded by the IF) is originally equal to about c$_i$, the speed of sound in ionized medium. This is highly supersonic w.r.t the sound speed $c_n(c_i/c_n \leq 10)$ in the neutral gas. The ionized gas sphere plays the role of a piston (fig. \ref{pic1}) and pushes a shock wave into the neutral gas. The shock then sets the neutral gas into motion outwards and it is therefore displaced from its original position.
The problem can be analyzed using the following  assumptions:

\begin{enumerate}
\item The layer of shocked neutral gas is thin because of the compression across the shock. A single radius R and velocity dR/dt then refers to both shock and IF.
\item The pressure behind the shock wave is assumed uniform in both the neutral and ionized gas.
\item The shock is strong, i.e., it moves highly supersonically into the neutral material.
\item The recombination rate in the ionized gas balances the stellar UV output rate.
\item The neutral gas ahead of the shock is at rest.  
\end{enumerate}

\begin{figure}
\begin{center}
\small
\psfrag{a}{n$_0$m$^{-3}$}
\psfrag{b}{C}
\psfrag{c}{R}
\psfrag{d}{$\delta$R$<<$R}
\psfrag{e}{S2}
\includegraphics[height=5cm]{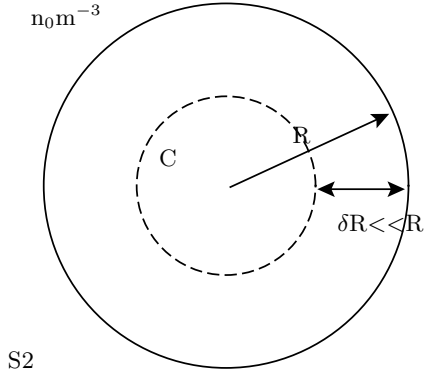}
\label{pic2}%
\caption{Interaction of Stellar Wind with ISM}
\end{center}
\end{figure}

Under these conditions, the pressure behind the shock wave, $P_s$, is equal to the pressure $P_i$ in the ionized gas given by
\begin{equation}
\label{eq12}
P_i=2n_ikT_i - n_im_Hc_i^2
\end{equation}

The pressure $P_s$ behind a strong shock is related to the shock velocity $V_s$ and density $n_0$ by assuming that the shock wave is isothermal.

\begin{equation}
\label{eq13}
P_s = r_oV_s^2
\end{equation}

Equation \ref{eq13} can be written as
\begin{equation}
\label{eq14}
P_s = P_i = n_om_H(dR/dt)^2
\end{equation}

where $dR/dt=V_s$. From equations \ref{eq12} and \ref{eq14}, neglecting thermal pressure, we get

\begin{equation}
\label{eq15}
\left(\frac {dR}{dt}\right)^2 =\left(\frac{n_i}{n_o}\right) c_i^2				
\end{equation}

Since the recombination rate in the ionized gas balances the stellar UV output rate outside Stromgren radius, using equation {\ref{eq4}, \ref{eq5} and \ref{eq15}, we get
\begin{equation}
\label{eq16}
R^{3/2}\left(\frac{dR}{dt}\right)^2= c_i^2 \left(\frac{3S^*}{4\pi n_o^2\beta_2}\right)^{1/2}  = c_i^2 R_s^{3/2}
\end{equation}

Taking the dimensionless variables,
\begin{equation}
\label{eq17a}
\lambda = R/R_s; N= c_i t/R_s
\end{equation}

Equation \ref{eq16} can be written as
\begin{equation}
\label{eq17}
\lambda^{3/4}\left(\frac{d\lambda}{dN}\right)=1
\end{equation}

The boundary condition to solve \ref{eq17} , is that the time taken to set up the initial Stromgren sphere is a very small fraction of the lifetime of the H$_{II}$ region, that is N=0 at $\lambda$=1.
Solution for equation \ref{eq17} is 

\begin{equation}
\label{eq19}
\lambda =\left(1+\frac{7N}{4}\right)^{4/7}
\end{equation}

\begin{equation}
\label{eq20}
\frac{d\lambda}{dN} =\left(1+ \frac{7N}{4}\right)^{-3/7}
\end{equation}

\noindent Equation \ref{eq20} implies that $dR/dt$ = $c_i$  at N = 0. 

The condition of final pressure equilibrium is when the hot ionized gas reaches 
pressure equilibrium with the surrounding cool neutral gas, that is 

\begin{equation}
\label{eq21}
2n_fkT_i=n_okT_n
\end{equation}

$N_f$ =ionized gas density at this stage.
$T_i$ and $T_N$ are respectively the ionized and neutral  gas temperatures.
The ionized gas sphere must still absorb all the stellar UV photons. Thus taking the clue from the equation \ref{eq4}, we get,

\begin{equation}
\label{eq22}
S^*=(4/3)\pi R_f^3n_f^2\beta_2
\end{equation}

R$_f$ =final radius of ionized gas sphere.

From equations \ref{eq5}, \ref{eq21} and \ref{eq22}, we get
\begin{equation}
\label{eq23}
(n_f/n_o)=(T_n /2T_i);  
R_f =(2T_i/T_n)^{2/3}R_s
\end{equation}

The ratio of fully ionized gas (M$_f$) to the total gas (M$_s$) contained within the initial Stromgren sphere of radius Rs is 

\begin{equation}
\label{eq24}
M_f/M_s=(R_f/R_s)^3(n_f/n_o)= 2T_i/T_n
\end{equation}

Typically, $n_f/n_0 \approx 0.005, R_f/ R_s\approx 34, M_f/ M_s \approx 200$. Hence at equilibrium we have,

\begin{equation}
\label{eq25}
\lambda = R / Rs \approx 34
\end{equation}

Equation \ref{eq19} then gives corresponding value of N at about 273 and therefore t $\approx$ 273 R$_s$/c$_i$.

It is interesting that since t$_{eqm}$ is about 8 10$^7$ years, typically an order of magnitude greater than phase of existence of massive star, it is unlikely that the final pressure equilibrium configuration can be reached unless n$_0$ is extremely high ($n_0 \geq 3~10^9 m^{-3}$)

\subsection{Interaction of stellar wind with ISM}

The sun emits a lot of gas in the form of a solar wind. The wind velocity ($V_{sw}~\approx$ 400 km/s) pushes the interstellar gas (where the sound speed is c$_i \approx$  10km/s) at a speed which is highly supersonic. The wind acts as a piston and a shock wave must immediately be set up in the interstellar gas. In doing this, the wind itself must be slowed down. To convert the kinetic energy of the wind to its thermal energy, we introduce a second shock into the wind itself. There are four distinct regions of flow and three distinct boundaries as marked in fig \ref{pic2}.

Region (a) is occupied by un-shocked stellar wind gas moving with a velocity $V_{sw}$ ($\approx$ 400 km/s). It enters the shock S1 that converts part of the energy of this wind into thermal energy. Region (b) contains shocked stellar wind gas. Since the Mach number of the shock is very high, this shock is very strong (see e.g. Harwit, 1998). The immediate post-shock gas temperature is 

\begin{equation}
\label{eq26}
T_s = \frac{3m_HV_{sw}^{2}}{32k} \approx 4~10^5 K
\end{equation}

It is the expansion of this hot bubble of gas, which drives the shell of shocked interstellar gas (region(c)), and this is sometimes referred to as an energy driven flow. The expansion velocity of the bubble drops because of the loss of internal energy as it does work on the surrounding material. This velocity very quickly drops to a value much less than the internal sound speed. 

Region (c) is a shell of interstellar gas, which has passed through the outer shock S2. Radiative cooling is now extremely effective in this region. Since the shock is a strong shock, the upstream to down stream velocity ration (shock strength) is about 4. Using the continuity condition, we then get the density immediately behind shock S2 to be  four times the ambient density. The interstellar gas enters shock S2 at a much lower speed than the velocity at which the stellar wind gas enters shock S1. The immediate post-shock temperature is therefore considerably lower than that behind S1. The gas behind S2 contains ions (such as O$^{++}$) which can be collisionally excited by electrons to produce the forbidden lines. Cooling is therefore very effective. Region (d) is the surrounding ionized interstellar gas with sound speed (10 km/s). Region (c) will be assumed to be a very thin shell and for the outer radius of the bubble of hot shocked wind gas. Further, since region (b) is thick, we will assume that the shocked wind gas occupies the entire volume interior to the shock surface of radius $R$. Since the interstellar gas is at rest, $dR/dt$ represents both the shock velocity relative to the interstellar gas and the expansion velocity of the bubble. Physically the thin shell (region(c)) is compressed and accelerated by the hot bubble, which has uniform pressure P.

\begin{figure}
\begin{center}
\small
\psfrag{a}{(a)}
\psfrag{b}{(b)}
\psfrag{c}{(c)}
\psfrag{d}{(d)}
\psfrag{e}{S1}
\psfrag{f}{S2}
\includegraphics[height=3cm]{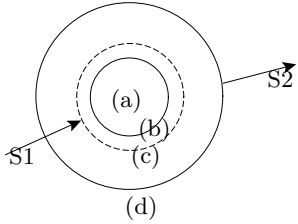}
\label{pic3}%
\caption{Regions of flow pattern}
\end{center}
\end{figure}
The equation of conservation of momentum of the shell has the form												
\begin{equation}
\label{eq27}
\frac{d}{dt}\left[\frac{4}{3}\pi R^3n_om_H\frac{dR}{dt}\right]=4\pi R^2P
\end{equation}
The left hand side of the equation \ref{eq27} represents the case of change of momentum of the shell, while the right hand side represents the force acting on it. Equation \ref{eq22} can be written in the form

\begin{equation}
\label{eq28}
 P=\frac{1}{3} \rho_o \frac{dR}{dt}^2+\left(\rho_o\frac{dR^2}{dt^2}R\right)
\end{equation}

The thermal energy per unit volume of a mono-atomic gas is equal to 3/2 times the gas pressure. The total thermal energy in region (b) therefore equals $4/3\pi R^3(3/2)P$. Conservation of energy for this region  is given as:                 

\begin{equation}
\label{eq29}
\frac{d}{dt}\left(\frac{3P}{2}4\pi R^3\right)=\frac{dE^*}{dt}-P \frac{d}{dt}(\frac{4}{3}\pi R^3)
\end{equation}

Equation \ref{eq29} states that the rate of change of thermal energy of region (b) (figure {\ref{pic3}}is equal to the rate at which the energy is fed into the region by gas entering shock S1 minus the rate at which the hot gas does work on the interstellar gas. This assumes that all the kinetic energy of the gas entering shock S1 is converted immediately into thermal energy. Equations 25 and 26 can be combined to give the equation of motion of the shell as
$$
\noindent R^4\left(\frac{dR^3}{dt^3}\right)+12R^3\left(\frac{dR}{dt}\right)\left(\frac{dR^2}{dt^2}\right)+15R^2\left(\frac{dR}{dt}\right)^3
$$
\begin{equation}
\label{eq30}
\indent = \frac{3}{2\pi\rho_o}\frac{dE^*}{dt}
\end{equation}
We assume the solution to be 
\begin{equation}
\label{eq31}
R=At^{\alpha}
\end{equation}

where A and $\alpha$ are positive constants. $R$ then increases with time and tends to zero at $t$=0. Inserting equation \ref{eq30} in equation \ref{eq31}, we get

$$ 
\noindent A^5[\alpha(\alpha-1)(\alpha-2)+12\alpha^2(\alpha-1)+15\alpha^3]t^{(5\alpha-3)} \\
$$
\begin{equation}
\label{eq32}
\indent =\frac{3}{2\pi\rho_o} \left(dE^*/dt\right)
\end{equation}

Since the mechanical output rate of the star is assumed constant in time both the R.H.S. and L.H.S of (29) are time independent and hence $\alpha$ = 3/5. It immediately follows that
\begin{equation}
\label{eq33}
A=\left(\frac{125}{154\pi}\right)^{1/5}\left(\frac{dE^*/dt}{\rho_o} \right)^{1/5}
\end{equation}

Therefore, we find
\begin{equation}
\label{eq34}
R = \left(\frac{125}{154\pi}\right)^{1/5}\left(\frac{dE^*/dt}{\rho_o}\right)^{1/5}t^{3/5}
\end{equation}

\begin{equation}
\label{eq35}
\frac{dR}{dt}=3\left(\frac{125}{154\pi}\right)^{1/5}\left(\frac{1}{\rho_o}\frac{dE^*}{dt}\right)^{1/5}t^{-2/5}
\end{equation}

Clearly, $R$ and ($dR/dt$) are related by
\begin{equation}
\label{eq36}
R=\frac{5t}{3}(\frac{dR}{dt})
\end{equation}

In the equilibrium condition, when the pressure of the solar wind is balanced by the ISM pressure, the condition can be specified as
\begin{equation}
\label{eq37}
P_{SW} = P_{ISM}
\end{equation}

The SW input pressure is $E^*$. Assuming adiabatic expansion the pressure falls as (see e.g. equation 11.31 Longair (1997)).

\begin{equation}
\label{eq38}
E^*(r) = E^*(0)\left(\frac{R}{R_s}\right)^{4/3}
\end{equation}
 
Combining equations 33 and 34 and assuming PISM = 1.5 n$_{ism} kT_{ism}$, at equilibrium,we get
\begin{equation}
\label{eq39}
1.5 n_{sw} kT_{sw}(R/R_s)^{4/3} =1.5 n_{ism} kT_{ism}
\end{equation}

Therefore,

\begin{equation}
\label{eq40}
 R=\left(\frac{n_{sw}T_{sw}}{n_{ism} T_{ism}}\right)^{3/4}R_s 
\end{equation}
Where $R_s$ = Radius of the Sun.

\subsection{Modulation of cosmic rays}

	Cosmic rays are charged particles that come to us from the galaxy. The interaction region between the ISM and the Sun heavily modulates them. We assume that the generic interstellar spectrum of the cosmic rays is given by the function of the form

\begin{equation}
\label{eq41}
f(E) = C E^{-2.5}
\end{equation}

\noindent where $E$ is the kinetic energy of the cosmic rays. This spectral form is assumed based on very high energy ($ \geq ~10^{14}$ eV) cosmic ray data. As we show below, these cosmic rays are not significantly modified by the various processes considered here. 

Since these are charged particles, they are significantly affected by the potential difference produced due to the interactions discussed above. The potential of interactions is empirically defined by

\begin{equation}
\label{eq42}
\phi = \int^R_{r(E)} (e C v_{sw}/k_d) dr 
\end{equation}
where C is the Compton Getting factor, e the electron charge $k_d$ is the diffusion coefficient, is given by
\begin{equation}
\label{eq43}
k = \frac{0.33 v_{cr}\lambda}{c} 
\end{equation}
and 
\begin{equation}
\label{eq44}
\lambda = 1/(n \pi \sqrt{2} d^2)
\end{equation}

where $\lambda$ is the mean free path, $d$ is the atomic diameter (\~ 10$^{-11}$ m) and n is the density of the medium. (see e.g. Lockwood and Webber, 1995, Garcia Munoz 1975, Garcia Munoz et al., 1990).  The differential flux of cosmic rays at any energy E of cosmic rays is given by 

\begin{equation}
\label{eq45}
J(E,f)= \frac{AE(E+2E_o)(E+\phi+m)^{-\alpha}}{(E+\phi)(E+2E_O+\phi)}
\end{equation}

\noindent where A, B, C, D are constants and equal to 5 108,1150, 650 and 1500 respectively. We obtain their values from the current best fits for cosmic ray spectra. $E$ is the kinetic energy in MeV/n, E$_0$ is the rest mass energy, and a is the slope  of the incident particles, m=B-Ce$^{(-E/D)}$.

\section{Calculations of the interaction parameters for the Sun under varying conditions of the ISM}

There is only one free parameter the Solar UV flux ($S^*$). For the present calculation, we take the UV flux to be 10$^{31}$ photons/sec (see e.g. Harwit, 1998, Stix, 1989) as the mean flux below 90 nm (13.6 eV) to evaluate the ionising strength of the Solar UV radiation. 

We calculate the sizes of the ionization sphere, the ionization pressure equilibrium sphere and the kinetic sphere produced around the sun due to different conditions of the interstellar medium. We define the various parameter values for the calculations of the interaction regions under the different conditions in table {\ref{SMtable}. We define six specific ISM environments as characteristic of regions encountered by the sun during the last 40 million years. Table \ref{ISMs} gives the values of the constants used and Tables \ref{SR}, \ref{IF}, and \ref{KE} the derived values of the parameters in the interaction regions. We define 2 types of fluff regions (fluff and fluff2). The parameters of fluff are taken from the observationally defined values (Frisch, 1998) and Fluff2 defines the parameters if the sun were in equilibrium with the fluff. The time scale for equilibrium is large and hence we use only the fluff as representing the current situation. The difference between the two versions is the mean density and temperature of the LISM.

\begin{table*}
\caption{Physical Parameters of different ISM conditions}
\label{ISMs}
\footnotesize
\begin{tabular}{lcccccc}
\hline
Parameter &		Fluff &	Fluff2	& Bubble &	Inter arm	& GMC	& Arm	 \\
\hline
n  (p/m$^3$) &		8.0 10$^{4}$ &	1.6 10$^{7}$	& 4.0 10$^{4}$ &	5.0 10$^{2}$ &	1.0 10$^{10} $ &	1.0 10$^{4}$	\\
T$_e$ (Kelvin)	&	7.0 10$^3$	& 1.0 10$^{3}$	& 1.0 10$^{6}$	& 1.0 10$^{2}$	& 1.0 10$^{1}$	& 1.0 10$^{3}$	\\
Time (seconds) &		1.0 10$^{2}$	& 1.0 10$^{2}$ &	1.0 10$^{5}$ &	1.0 10$^{7}$ &	1.0 10$^{8}$ & 	1.010$^{9}$	\\
Recomb coeff. $\beta_2$ (m$^3$/s)	& 2.61 10$^{-19}$ &	1.12 10$^{-18}$ &	6.32 10$^{-21}$ &	6.32 10$^{-18}$ &	3.56 10$^{-17}$ &	1.1210$^{-18}$ \\
Recom Fact $\alpha$ &	6.22 10$^{-13}$ &	1.64 10$^{-12}$ &	5.20 10$^{-14}$	& 5.20 10$^{-12}$	& 1.64 10$^{-11}$ &	1.64 10$^{-12}$ \\
Recomb time t$_R$ (seconds) & 	4.78 10$^{13}$	& 1.11 10$^{13}$	& 1.98 10$^{15}$ &	1.98 10$^{12}$ &	3.51 10$^{11}$ &	1.11 10$^{13}$	 \\
M$ism$	(Kg)  &	1.86 10$^{21}$ &	3.73 10$^{23}$ &	9.32 10$^{20}$	& 1.16 10$^{19}$	& 2.33 10$^{26}$ &	2.33 10$^{20}$ \\
V$_{sound} (m/s) $	& 1.21 10$^{4}$ &	4.56 10$^{3}$ &	1.44 10$^{5}$ &	1.44 10$^{3}$ & 	4.56 10$^{2}$ &	4.56 10$^{3}$ \\
\hline
\end{tabular}
\end{table*}
\begin{table*}
\caption{Interaction parameters for Stromgren radius}
\label{SR}
\footnotesize
\begin{tabular}{lcccccc}
\hline
Parameter &		Fluff: &	Fluff2	& Bubble &	Inter arm	& GMC	& Arm	 \\
\hline
Stromgren radius R$_s$ (meters) & 	1.13 10$^{13}$ &	2.02 10$^{11}$ &	6.18 10$^{13}$ &	1.15 10$^{14}$ &	8.76 10$^{8}$ &	2.77 10$^{13}$ \\
Stromgren time T$_s$ (sec)&	7.38 10$^{8}$ &	3.51 10$^{7}$ &	3.39 10$^{8}$ &	6.29 10$^{10}$ &	1.52 10$^{6}$ &	4.80 10$^{9}$ \\
Stromgren vel V$_s$ (m/s) &	1.53 10$^{4}$ &	5.77 10$^{3}$	& 1.82 10$^{5}$ &	1.82 10$^{3}$	& 5.77 10$^{2}$ & 	5.77 10$^{3}$ \\
Stromgren Temp (K) &	2.10 10$^{4}$ &	2.10 10$^{4}$ &	2.10 10$^{4}$ &	2.10 10$^{4}$ &	2.10 10$^{4}$ &	2.10 10$^{4}$ \\
Density (/m$3$)&	2.67 10$^{4}$ &	7.62 10$^{5}$ &	1.90 10$^{6}$ &	2.38 10$^{0}$ &	4.76 10$^{6}$ &	4.76 10$^{2}$  \\
dR/dt (initial) (m/s) &	7.58  10$^{-2}$ &	1.21 10$^{0}$ &	5.21 10$^{-3}$ &	1.21 10$^{-1}$ &	1.04 10$^{2}$ &	1.04 10$^{-1}$ \\
dR/dt (final) (m/s) &	1.12 10$^{-15}$ 	& 5.55 10$^{-15}$ & 0.0  &	5.27 10$^{-16}$ &	9.81 10$^{-13}$ &	0.0  \\
Particle Velocity (m/s) &	2.29 10$^{4}$ &	2.29 10$^{4}$ &	2.29 10$^{4}$ &	2.29 10$^{4}$ &	2.29 10$^{4}$	& 2.29 10$^{4}$ \\
Mean free path (m) &	8.44 10$^{16}$ &	2.95 10$^{15}$ &	1.18 10$^{15}$ &	9.45 10$^{20}$	& 4.73 10$^{14}$ &	4.73 10$^{18}$ \\
Diffusion Coefficient (/m$^2$s,c=1) &	1.19 10$^{16}$ &	4.18 10$^{14}$ &	1.67 10$^{14}$ &	1.34 10$^{20}$ &	6.68 10$^{13}$ &	6.68 10$^{17}$ 	 \\
$\phi$	(MV) &	8.64 10$^{1}$ &	4.44 10$^{1}$ &	3.39 10$^{4}$ &	7.86 10$^{-2}$ &	1.20 10$^{0}$& 3.79 10$^{0}$\\
\hline
\end{tabular}
\end{table*}

\begin{table*}
\caption{Interaction parameters for Ionization equilibrium}
\label{IF}
\footnotesize
\begin{tabular}{lcccccc}
\hline
Parameter &		Fluff: &	Fluff2	& Bubble &	Inter arm	& GMC	& Arm	\\
\hline
R$_{ther}${IH} (R$_IH$ = 34R$_s$) (m) &	3.83 10$^{14}$ &	6.88 10$^{12}$ &	2.10 10$^{15}$ &	3.90 10$^{15}$ &	2.98 10$^{10}$ &	9.41 10$^{14}$	\\
t${_{thermal}}$IH (t$_{IH}$) (s) &	2.55 10$^{11}$ &	1.21 10$^{10}$ &	1.17 10$^{11}$ & 2.17 10$^{13}$ &	5.24 10$^{8}$ &	1.66 10$^{12}$ \\
V$_{IH}$ (m/s) &		1.50 10$^{3}$ &	5.68 10$^{2}$	& 1.80 10$^{4}$	& 1.80 10$^{2}$	& 5.68 10$^{1}$ &	5.68 10$^{2}$  \\
Temperature (K) &	7.00 10$^{5}$ &	1.00 10$^{5}$ &	1.00 10$^{8}$ &	1.00 10$^{4}$ &	1.00 10$^{3}$ &	1.00 10$^{5}$ \\
Density	(/m$^3$) &	4.00 10$^{2}$	& 8.00 10$^{4}$ &	2.00 10$^{2}$ &	2.50 10$^{0}$ &	5.00 10$^{7}$ &	5.00 10$^{1}$ 	\\
dR/dt (m/s) &		6.03 10$^{1}$	& 2.28 10$^{1}$ &	7.21 10$^{2}$ &	7.21 10$^{0}$ &	2.28 10$^{0}$	& 2.28 10$^{1}$  \\
Magnetic field (Tesla) &	1.50 10$^{-12}$ &1.50 10$^{-12}$ &1.50 10$^{-12}$ &1.50 10$^{-12}$ &1.50 10$^{-12}$ &1.50 10$^{-12}$ \\
Particle Velocity (m/s) &	1.32 10$^{5}$ &	4.99 10$^{4}$ &	1.58 10$^{6}$ &	1.58 10$^{4}$ &	4.99 10$^{3}$ &	4.99 10$^{4}$ \\
Mean free path  (m) &	5.63 10$^{18}$ &	2.81 10$^{16}$ &	1.13 10$^{19}$ &	9.00 10$^{20}$ &	4.50 10$^{13}$ &	4.50 10$^{19}$ \\
Diffusion Coefficient (/m$^2$s,c=1)  &	7.95 10$^{17}$ &	3.98 10$^{15}$ &	1.59 10$^{18}$ &	1.27 10$^{20}$ &	6.36 10$^{12}$ &	6.36 10$^{18}$	\\
$\phi$ (MV)	&	2.54 10$^{2}$ &	3.46 10$^{2}$ &	8.34 10$^{3}$ &	1.94 10$^{0}$ &	9.35 10$^{1}$ &	2.96 10$^{1}$ \\
\hline
\end{tabular}
\end{table*}

\begin{table*}
\caption{Interaction parameters for Heliopause}
\label{KE}
\footnotesize
\begin{tabular}{lcccccc}
\hline
Parameter &		Fluff: &	Fluff2	& Bubble &	Inter arm	& GMC	& Arm	\\
\hline
$R_{HP}$ (m) &		4.12 10$^{12}$ &	3.34 10$^{11}$ &	1.68 10$^{11}$ &	4.49 10$^{15}$ &	8.44 10$^{10}$	& 8.44 10$^{13}$ \\
$t_{HP}$	(s) &	3.42 10$^{8}$	& 7.32 10$^{7}$ &	1.16 10$^{6}$ &	3.11 10$^{12}$ &	1.85 10$^{8}$ &	1.85 10$^{10}$ \\
$V_{HP}$  (m/s)&	1.21 10$^{4}$	& 4.56 10$^{3}$ &	1.44 10$^{5}$ &	1.44 10$^{3}$ &	4.56 10$^{2}$ &	4.56 10$^{3}$  \\
Temperature	 (K) & 6.00 10$^{6}$ &	6.00 10$^{6}$ &	6.00 10$^{6}$ &	6.00 10$^{6}$ &	6.00 10$^{6}$ &	6.00 10$^{6}$ \\
Particle Density (/m$^3$) 	& 1.20 10$^{5}$ &	3.43 10$^{6}$ &	8.58 10$^{6}$ &	1.07 10$^{1}$ &	2.15 10$^{7}$ &	2.15 10$^{3}$ \\
Mean free path (m)	& 1.87 10$^{16}$ &	6.56 10$^{14}$ &	2.62 10$^{14}$ &	2.10 10$^{20}$ &	1.05 10$^{14}$ &	1.05 10$^{18}$ \\
Diffusion Coefficient	(/m$^2$s,c=1)	& 2.65 10$^{15}$ &	9.27 10$^{13}$ &	3.71 10$^{13}$ &	2.97 10$^{19}$ &	1.48 10$^{13}$ &	1.48 10$^{17}$ \\
$\phi$ (MV) &	2.49 10$^{3}$ &	5.76 10$^{3}$ &	7.24 10$^{3}$ &	2.42 10$^{2}$ &	9.11 10$^{3}$ &	9.11 10$^{2}$ \\
\hline
\end{tabular}
\end{table*}

The basic features of the calculations are given in figures \ref{fig1} and \ref{fig2}. The figures show the equilibrium interaction radius and the time scale required for reaching the equilibrium for the Stromgren Radius (SR), the Ionization Front (IF) and the Heliopause (HP). Both the axes are in log scales. In the history of the Sun (table \ref{SMtable}) indicates that the X-axis indicates time before present in the movement of the Sun in the Galaxy.

The consequences of this changing ISM environment can be quite varied. The present studies indicate, for example that when the solar system was passing through a high density cloud, the drag on meteors and potential instabilities could trigger a large number of events such as increased cometary showers, changes in the solar constant etc. since it is estimated that almost 98 \% of the interplanetary material is of ISM origin (Frisch, 1998). Also, all models of stellar evolution essentially assume that the environmental pressure on the star is negligible but clearly, we know that the Heliosphere size is significantly altered by a factor of 2 or larger even due to small changes in the solar activity (see e.g. Lockwood and Webber, 1995). The anomalous cosmic rays in the energy range of 10 to 100 MeV/n are particularly sensitive to the heliospheric condition. Since their origin is related to the ionisation of interstallar neutrals into the solar system (Cummings and Stone, 1998). Recently, several authors: Czechowski et al.,2001; Izmodenov, 2001, Donato, Manrin and Tailler, 2002; and references therein. See also Jokipii et al., 1995 and Jokipii, 1996) have attempted to use the warm neutral ions as well as galactic cosmic ray modulation to determine the local conditions. All these studies rely on measurements of either anomalous cosmic ray studies or on galactic cosmic rays in the energy range up to 10$^{10}$ eV to determine the current local ISM conditions. These are either restricted to deriving inferences of the Heliopause or on  time average studies of cosmic rays confined in the local bubble. Since the life time of cosmic rays is large (15 million years, see Mewaldt et al., 2001), such studies will not be able to give a time sequence of LISM in the past. For high energy cosmic rays however, the origin itself is in supernova shock fronts (see e.g. Lagage and Cesarsky, 1981) the gyro radius of higher energy cosmic rays may be more sensitive to the LISM conditions since their fluxes would be directly modulated by the LISM.

\section{Discussion}

\begin{figure}
\centering
\includegraphics[height=2.5in,width=4in]{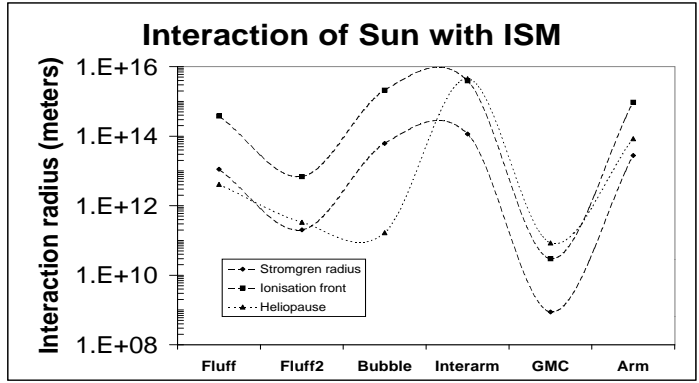}
\caption{Equilibrium interaction radius under different ISM conditions}
\label{fig1}%
\end{figure}

\begin{figure}
\centering
\includegraphics[height=2.5in,width=4in]{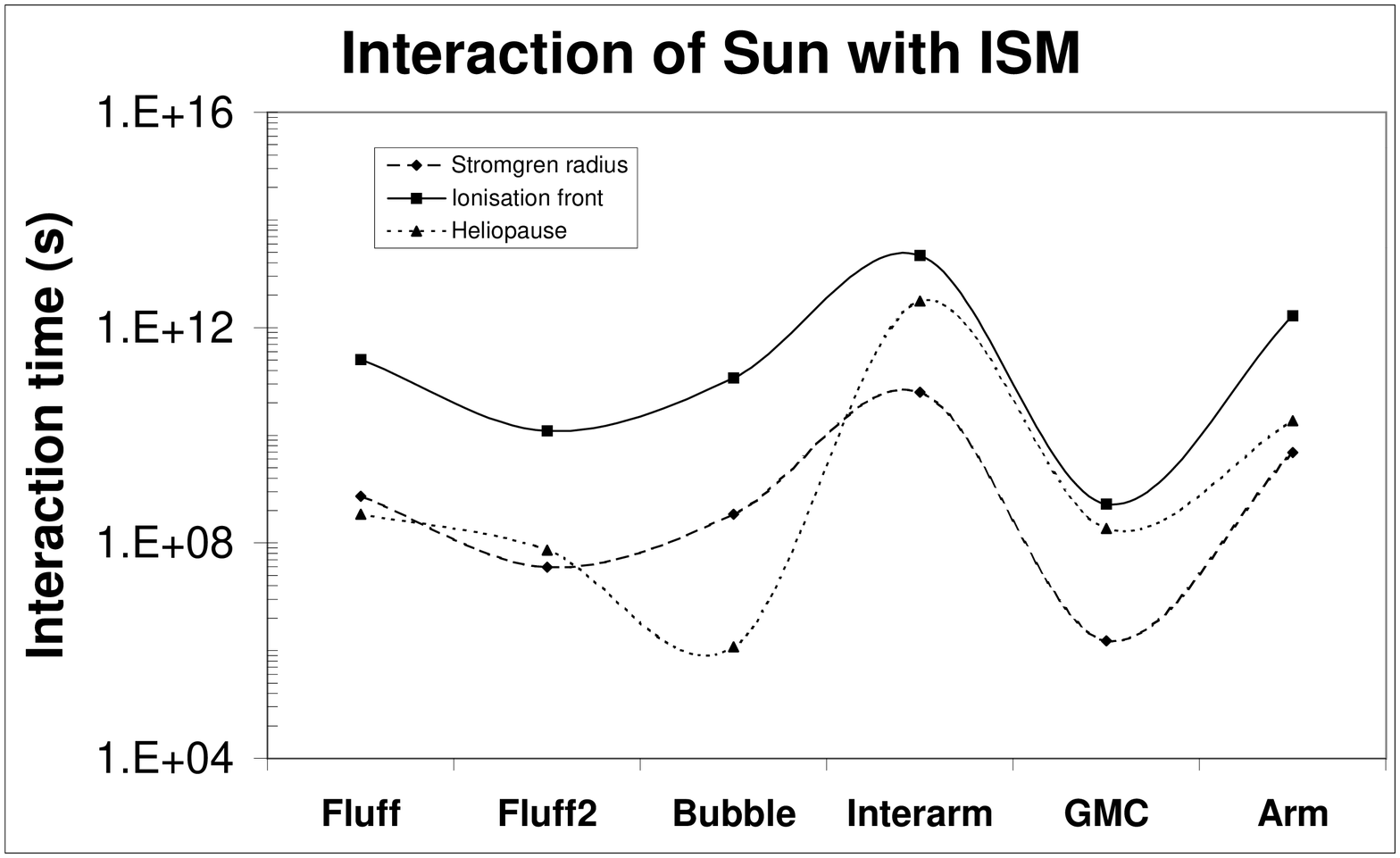}
\caption{Evolution of the ionization induced interaction regions}
\label{fig2}%
\end{figure}

We have displayed in figures \ref{fig1} and \ref{fig2} the depth of the interaction region under different ISM conditions and different physical processes. As can be seen from figure \ref{fig1} the SR is 4 orders of magnitude larger than the HP that is normally considered the boundary region of the solar interaction with the interstellar medium. While the SR decreased by about two orders of magnitude as the sun entered the local fluff inside the supernova, the HP remained more or less constant during the movement of the Sun from bubble to fluff. This is because the HP depends on the density and temperature of the medium while SR depends only on the ISM density. It should also be noted that the variations in all the three radii can be of the order of 6 orders of magnitude. The time scales to reach equilibrium are also quite varied (figure \ref{fig2}). As can be seen from the figure, in the low-density region of the bubble, the HP reaches its equilibrium in less than 1 year but it would take about 100 years to reach an equilibrium in the fluff. On the other hand, the local environment of the GMC, the SR and IF would not reach their equilibrium state for any realistic time scales. 

In order to look for possible direct, observational data to study and understand the changing LISM conditions over long time scales, we calculate the modulation of cosmic rays over these time scales based on the formulation given above.	

\begin{figure}
\centering
\includegraphics[height=2.5in,width=4in]{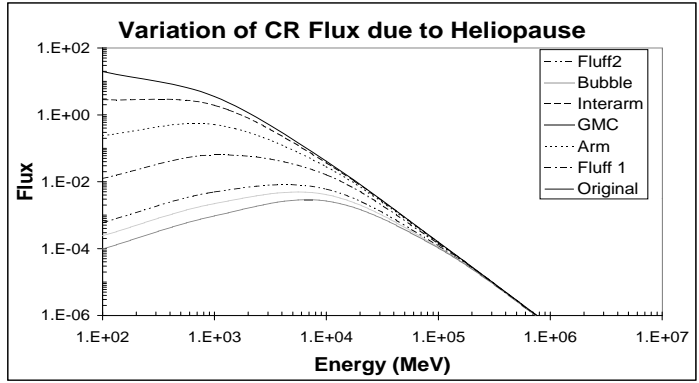}
\caption{Variation in the cosmic ray flux due to Heliopause}
\label{fig3}%
\end{figure}

\begin{figure}
\centering
\includegraphics[height=2.5in,width=4in]{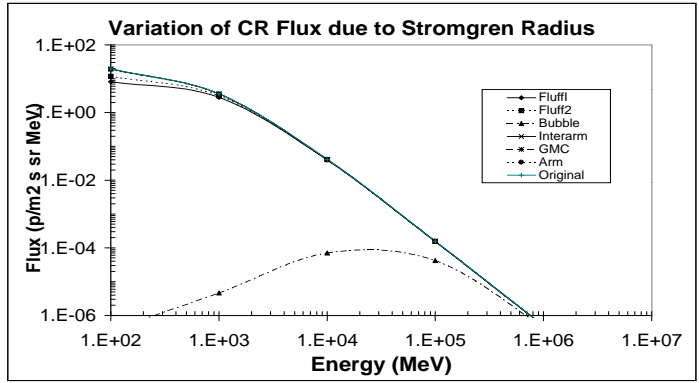}
\caption{Variation in the cosmic ray flux due to Stromgren Radius}
\label{fig4}%
\end{figure}
    
\begin{figure}
\centering
\includegraphics[height=2.5in,width=4in]{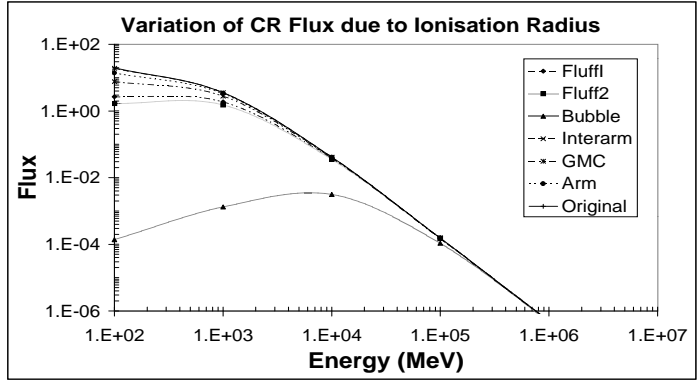}
\caption{Variation in cosmic ray flux due to Ionization front}
\label{fig5}%
\end{figure}

In figures \ref{fig3}, \ref{fig4} and \ref{fig5} we have plotted the cosmic ray modulation under different LISM conditions. In all cases we assume that the cosmic rays with a spectral index of -2.5 in energy (equation \ref{eq41} impinge on the sphere. We calculate the potential generated by each sphere and evaluate the flux parameters. This is the most dramatic of all variations except when the solar system enters the bubble. 
	
	 In figure \ref{fig3}, the variations due to the changes in the Heliopause alone are shown As can be seen from the figure, in the low energy can be as much as 6 orders of magnitude due to the variation of the HP alone. Note that a very high energy density in cosmic rays comes from the low energy anomalous cosmic rays, which are accelerated in the HP itself. For the present study we have ignored this component.
 
	In figure \ref{fig4} we show the effects of SR on the cosmic ray modulations. Since this shock front is generated due to ionization and since its radius is much larger, the modulation is also more dramatic. Figure \ref{fig5} shows the modulation produced by the IF which extends beyond SR and therefore also affects the CRs more severely. 

\begin{figure}
\centering
\includegraphics[height=2.5in,width=4in]{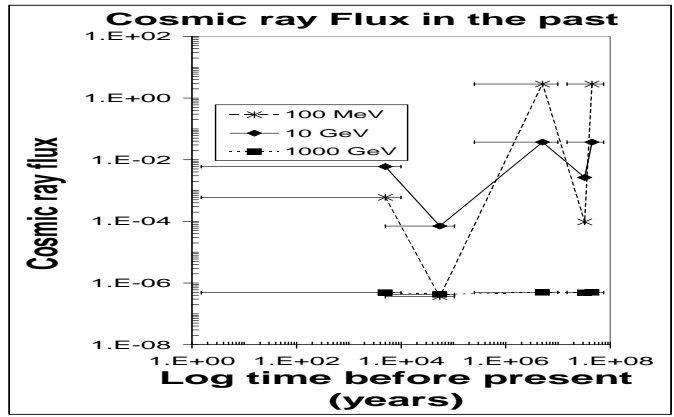}
\caption{Variation in cosmic ray flux due to variation in the LISM conditions}
\label{CRflux}%
\end{figure}

In figure{\ref{CRflux} we have plotted the cosmic ray energy density over a period of time. The flux is calculated based on the assumption that the solar system is embedded in a specific environment as given by table {\ref{SMtable}} and that all the three forms of interactions discussed above are in equilibrium. We plot the {\it {minimum}} flux that will pass through the three interaction regions and take that as the final flux that would be seen by the solar system. Note that the two phases of passage through a SN and an H$_{II}$ front are not plotted since these are period of local and intense cosmic ray re-acceleration which have been ignored in the present work. 	If the present analysis is borne out by further studies, this may be an important component in producing the knee in the cosmic ray spectra.

Florinski Zank and Pogorelov (2003), Florinski, Zank and Axford (2003) and Florinski and Jokipii (2003) have used similar transport calculations to discuss the affects of dense molecular cloud on cosmic ray propagation both, for galactic cosmic rays and for anomalous cosmic rays. Their results show that cosmic rays would be significantly modulated in the Heliosphere (Florinski and Jokipii, 2003). The study reported in Florinski, Zank and Pogorelov (2003) show work out the modulation formulation for cosmic rays under such environment while Florinski, Zank and Axford (2003) show that the $^10$Be data can best be explained by assuming that the solar system passed through a molecular cloud 35,000 to 60,000 years before present.

\section{Conclusions}

We have investigated the interaction of the Sun with the local interstellar medium. We show that there have been dramatic changes in the LSIM condition in the last 40 million years or longer. We then investigate the physical processes through which the Sun interacts with the LISM and show how these regions of interaction would be dramatically different for varying ISM conditions. We investigate radiative modulation, radiation induced ionization front and the classical solar wind interaction of the ISM. We then investigate the affects of these interaction regions on the modulation of cosmic rays. We find that the first two parameters will have significant affect on cosmic ray modulation though this has generally been neglected in the literature.

\section{acknowledgements}

The author wishes to thank Prof. A. W. Wolfendale for his assistance and encouragement during all stages of this work and his careful evaluation of the manuscript. The author wishes to acknowledge the efforts taken by Prof. S M Chitre is commenting on the manuscript and giving his valuable comments which have gone a long way in improving this manuscript. The author also wishes to thank Prof. D Lal for his incessant questioning of the problem which considerably improved our insights to the problem. The author also wishes to thank Ms. S. Nair for her help in the analysis of data. 

\section{References}
\begin{itemize}
\item	Bash F, 1986, The Galaxy and the Solar System, ed: R. Smoluchowski, Bahcall J N III, Shapley M M, University of Arizona Press, pp 35
\item Breitschwerdt, D, 2001, Ap\&SS, 276, 163
\item	Breitschwerdt, D, 1998, Proceedings of the 166 IAU Symposium on ''The local bubble and beyond'',  Lecture Notes in Physics, Springer Verlag, ed D. Breitschwerdt, M Freyberg, J Trumper,
\item	Cluber S V M, Napier W M, 1984, MNRAS, 208, 575
\item	Cummings A C and Stone E C, 1996, SSR, 78, 117
\item	Cummings A C, Mewaldt R A, Blake J B, Cummings J R, Franz M, Hovestadt D, Klecker B, Mason G M, Mazur J E, Stone E C, von Rosenvinge T T, Webber W R, 1995, GRL, 22, 341
\item Cummings A C, Stone, E C, 1998, SSRv, 83, 51
\item Czechowski, A, Fichtner II, Grzedzielski S, Hilchenbach, M, Ilsich, K C, Jokipii J R, Kansch T, Kota J and Shaw, A, 2001, A\&A, 622, 634
\item Dehnen W, Binney J J, 1998, MNRAS, 298, 386
\item Donato F, Manrin D, Taillet R, 2002, A\&A, 381, 538
\item	Dyson J E and Williams D A, 1997, The Physics of the Interstellar Medium, Institute of Physics, UK
\item	Egger R J, Freyberg M J, Morfill G E, 1996, SSR, 75, 511
\item Florinski V, Zank G P and Axford W L, 2003, GRL, 30, 2206
\item Florinski V, Zank G P and Pogorelov N V, 2003, 108, no A6, 1228 (pp SSH 1)
\item Florinski V and Jokipii J R, 2003, ApJ, 591, 454
\item Franco, G A P, 2002, MNRAS, 331, 474
\item	Frisch P C, 1995, SSR, 72, 499
\item Frisch, P C, 1998, SSR, 86, 107
\item	Fujii Z, McDOnald F B, JGR, 1997, 102, 24201
\item	Garcia Munoz et al., 1975, ApJ, 202, 265
\item	Garcia-Munoz M, Pyle K R, Simpson J A,  1990, XXI ICRC, Adiled
\item	Harwit M, 1998, Astrophysical Concepts, John Wiley and Sons
\item Haffner L M, 2002, AAS, 200, 3309
\item Ingrid M, Hiroshi, K, 2001, SSR, 97, 389
\item Izmodenov V, Gruntman M, Baranov V, Fahr H, 2001, SSRv, 97, 413
\item	Jokipii J R, 1996, ApJ, 644, L47
\item	Jokipii J R, Kota J, Merenyi, E, 1993, ApJ, 405, 782
\item	Karmesin S R, Liewer P C, Brackbill, J U, 1995, GRL, 22, 1153
\item Lagage, P. O. Cesarsky, C. J., 1981, A\&A, 125. 249
\item	Lockwood J A and Webber W R, 1995, ApJ, 442, 852
\item	Longair M S, 1992, High Energy Astrophysics, Volume 1 and 2, Cambridge University Press
\item Maiz-Apellaniz, J, 2001, ApJ, 560, L83
\item Meisel D D, Diego J and Mathews J D, 2002, ApJ, 567, 323
\item Mewaldt R A, Yanasak, N E, Wiedenbeck M E, Davis A J, Binns W R, Christian E R, Cummings A C, Hink P, L, Leske R A, Niebur S M, Stone E C, Von Rosenvinge T T, 2001, SSRv, 99, 27
\item Moos H W, et al., 2002, ApJS, 140,3
\item	Nagahama T, Akira M, Hideo O, Yasuo F, 1998, AJ, 116, 336
\item	Napier W M, 1985, Dissipation of a primordial cloud of comets: Their origin and evolution, IAU Colloquium 83, ed. A Carusi and G B Valsecchi, Dordrecht: D Reidel), pp 41
\item Pavlov G G, Welty A D, Cordova, F A, 1997, ApJ, 489, L75, 1997
\item	Poppel W, 1997, Fundamentals of Physics, 18, 1
\item	Ramadurai S, 1993, BASI, 21, 391, 1993
\item Ramesh Bhat, N D, Gupta Y, Pramesh Rao A, 2001, ApSS, 276, 227
\item Ramesh Bhat, N D, Gupta Y, 2002, ApJ, 567, 342
\item Redfield S, Linsky J L, 2002, ApJS, 139, 439
\item Ricardo G, Beckman, J E, 2001, ApSS, 276, 187
\item Rothmann G J, Woods T N, White O R and London J, 1994, in Sun as a variable Star, Proceedings of IAU Colloquium 143, ed Pap J M, Frohilch C, Hudson H SSolanki S K, Cambridge Univ Press, page 73
\item Seth R, Linsky J L, 2001, ApJ, 551, 413
\item Shelton, R L, 2002, ApJ, 569, 758
\item Smith R K, Cox D P, 2001, ApJS, 134, 283
\item Smoker J V, Keenana, F P, Lehner, N, Trundle, C, 2002, A\&A,, 387, 1057
\item Stix M, 1989, {\it {The Sun}}, Springer Verlag
\item	Szabelska B, Szabelski J, Wolfendale A W, 1991, JP(G), 17, 545
\item Vahia M N, Lal D, 2002, {\it {Origin of elements in the Solar System}}, ed O Manuel, Kluwer Press, 351
\item	Van der Walt D J, Wolfendale A W, 1988, JP(G), 14, L159
\item	Whiteoak J B Z, Green A J, 1996, A\&A, 118, 329
\item	Wichmann R, Sterzik M, Krautter J, Metanomski A, Voges W, 1997, A\&A, 326, 211
\end{itemize}
\end{document}